\documentclass[letter]{aa} 

\usepackage{graphicx}

\usepackage{txfonts}
\begin{document}

   \title{The impact of corotation on gradual solar energetic particle event intensity profiles
   }

   \subtitle{}

   \author{A. Hutchinson
          \inst{1}
          \and
          S. Dalla\inst{1}
          \and
          T. Laitinen\inst{1}
          \and
          C. O. G. Waterfall\inst{1}}

   \institute{Jeremiah Horrocks Institute, University of Central Lancashire, Preston, PR1 2HE, UK\\
                \email{AHutchinson3@uclan.ac.uk}}
             

 
  \abstract
   {Corotation of particle-filled magnetic flux tubes is generally thought to have a minor influence on the time-intensity profiles of gradual Solar Energetic Particle (SEP) events. For this reason many SEP models solve the focussed transport equation within the corotating frame, thus neglecting corotation effects.}
   {We use simulations to study the effects of corotation on gradual SEP intensity profiles at a range of observer longitudinal positions relative to the solar source. We study how corotation affects the duration and decay time constant of SEP events and the variation of peak intensity with observer position.}
   {We use a 3D full-orbit test particle code with time-extended SEP injection via a shock-like source. Unlike with focussed transport models, the test particle approach enables us to switch corotation on and off easily. While shock acceleration and downstream features are not modelled directly, our methodology allows us to study how corotation and the time-varying observer-shock magnetic connection influence intensity profiles detected at several observers.}
   {We find that corotation strongly affects SEP intensity profiles, for a monoenergetic population of 5 MeV protons, being a dominant influence during the decay phase. Simulations including corotation display dramatically shortened durations for western events, compared to those which do not include it. When corotation effects are taken into account, for both eastern and western events the decay time constant is reduced and its dependence on the value of the scattering mean free path becomes negligible. Corotation reduces the SEP peak intensity for western events and enhances it for eastern ones, thus making the east-west asymmetry in peak intensity stronger, compared to the no-corotation case. Modelling SEP intensity profiles without carefully accounting for the effects of corotation leads to artificially extended decay phases during western events, leading to profiles with a similar shape regardless of observer longitudinal position.}
  {}

   \maketitle
%

\section{Introduction}

Solar energetic particles (SEPs) in the $\sim$1-10 MeV/nuc range in large gradual events are thought to be accelerated at Coronal Mass Ejection (CME)-driven shocks as they propagate throughout the corona and interplanetary space. The characteristic long duration decays measured by spacecraft at 1 au are assumed to arise from acceleration at the shock that is extended in time \citep{cane_1988,Reames_1997,Desai2016}.

Magnetic flux tubes that guide energetic particle propagation corotate with the Sun. Depending on the location of an observing spacecraft, corotation may carry SEP-filled flux tubes either towards or away from the observer. Several authors have pointed out the importance of this effect on SEP intensity profiles for the case of an instantaneous injection at the Sun \citep{Droge_2010,Giacalone_2012,Marsh_2015,Laitinen_2018}. Corotation of steady state and quasi-steady state solar wind structures is the basis of empirical solar wind forecast models \citep[e.g.][]{owens_2013}. 

Relatively few studies have commented on the effects of corotation on SEP intensity profiles for the case of time-extended injection at a CME-driven shock. Using an approximate methodology for including corotation within a focussed transport model, \cite{Kal&Wib_1997} concluded that it is not very important for events with an injection that is extended in time. 
\cite{Lario_1998} also studied the effect of corotation in a focussed transport model by considering particle injection into a discrete number of flux tubes that sequentially pass over the observer. They noted that for certain periods corotation could affect their derived injection rate by up to a factor of 1.4. However, they concluded that corotation is not relevant in most situations within 1 au.
Overall, corotation is not thought to play a major role in large, gradual SEP events. Possibly as a result, many studies modelling extended SEP shock-like injections solve the relevant equations (usually focussed transport equations) in the corotating frame, thus neglecting corotation effects \citep[e.g.][]{ Wang_2012, He&Wan_2017, Hu_2018}.


Although corotation is largely neglected it is natural to assume it may play a role in the east-west asymmetries found for a number of SEP intensity profile parameters. Several studies of SEP measurements have reported east-west asymmetries.
In their study of 35 SEP events \cite{Lario_2013} analysed the longitudinal dependence of peak intensities, finding that the distribution is centred at eastern longitudes. 
This finding was confirmed by \cite{Richardson2014}, who also found east-west asymmetries in proton onset delays and time to peak intensity. 


\cite{Ding_2022} investigated the east-west asymmetry in time-averaged fluence and peak intensity by studying multi-spacecraft SEP events and modelling the SEP fluence at 10 observers using their 2D iPATH model. They concluded that the observed east-west asymmetric distribution can be explained by the combined effect of time-extended shock acceleration  and the geometry of Interplanetary Magnetic Field (IMF) lines, as well as determining that slower solar wind speeds and faster CME speeds enhance the east-west asymmetry. Although they stated that corotation is included in their model, its effect on observables was not discussed.


\cite{He_2015} used a 5D focussed transport equation to model SEP propagation from several solar sources separated by $30^\circ$ longitude. Comparing cases where the separation between the source and the observer's magnetic footpoint was the same, they found that eastern sources relative to the observer produced systematically larger SEP peak fluxes. They concluded that the east-west azimuthal asymmetry in the IMF and the effects of perpendicular transport lead to the longitudinally asymmetric distribution of SEPs.
This conclusion was supported by \cite{He&Wan_2017} who investigated 78 solar proton events finding that, for the same flare-footpoint separation, the number of solar proton events is larger for eastern solar sources.

An alternative methodology to focussed transport approaches for describing SEP propagation through interplanetary space consists of using 3D test particle simulations, in which the trajectories of an SEP population of specified initial parameters can be calculated  \citep[e.g.,][]{Marsh_2013,dalla_2020,Waterfall_2022}. While focussing on different aspects of SEP propagation, the latter studies have all considered instantaneous injections taking place close to the Sun. The results of the simulations \citep[e.g.][]{Marsh_2015,dalla_2017,dalla_2020} produce time-intensity profiles with characteristic ordering by east-west observer longitudes \citep{cane_1988}.

In this paper we study the role of corotation effects in SEP events by means of 3D test particle simulations, with time-extended particle injection, describing continuous acceleration at a CME shock-like source. The test particle approach provides a natural way to describe corotation via the presence of a solar wind electric field, as described below. It is also easy to remove corotation by switching off the electric field.

Rather than a full shock model, we use a concentrically expanding moving shock-like source as an initial approximation. Features of the shock's downstream region are not modelled. The latter is often characterised by a complicated magnetic configuration, due to the presence of a flux rope \citep[e.g.][]{Zurbuchen2006} and non-Parker magnetic field lines \citep[see e.g.][Figure 1]{Lario_1998}. A moving shock model similar to ours has previously been used by \cite{Kal&Wib_1997} and \cite{Wang_2012}.

Our simulations show that although the role of corotation is generally ignored in the interpretation of gradual SEP events, it plays a major role in shaping observables, and can impact the east-west asymmetries in a variety of parameters.
The layout of the paper is as follows: in section \ref{sec:sims} we describe our simulations, in section \ref{sec:int_obs} we consider the effect of corotation on intensity profiles, and the discussion and conclusions can be found in section \ref{sec:disc}.

\begin{figure*}[htb]
    \centering
    \includegraphics[width = 0.9\linewidth, keepaspectratio = true]{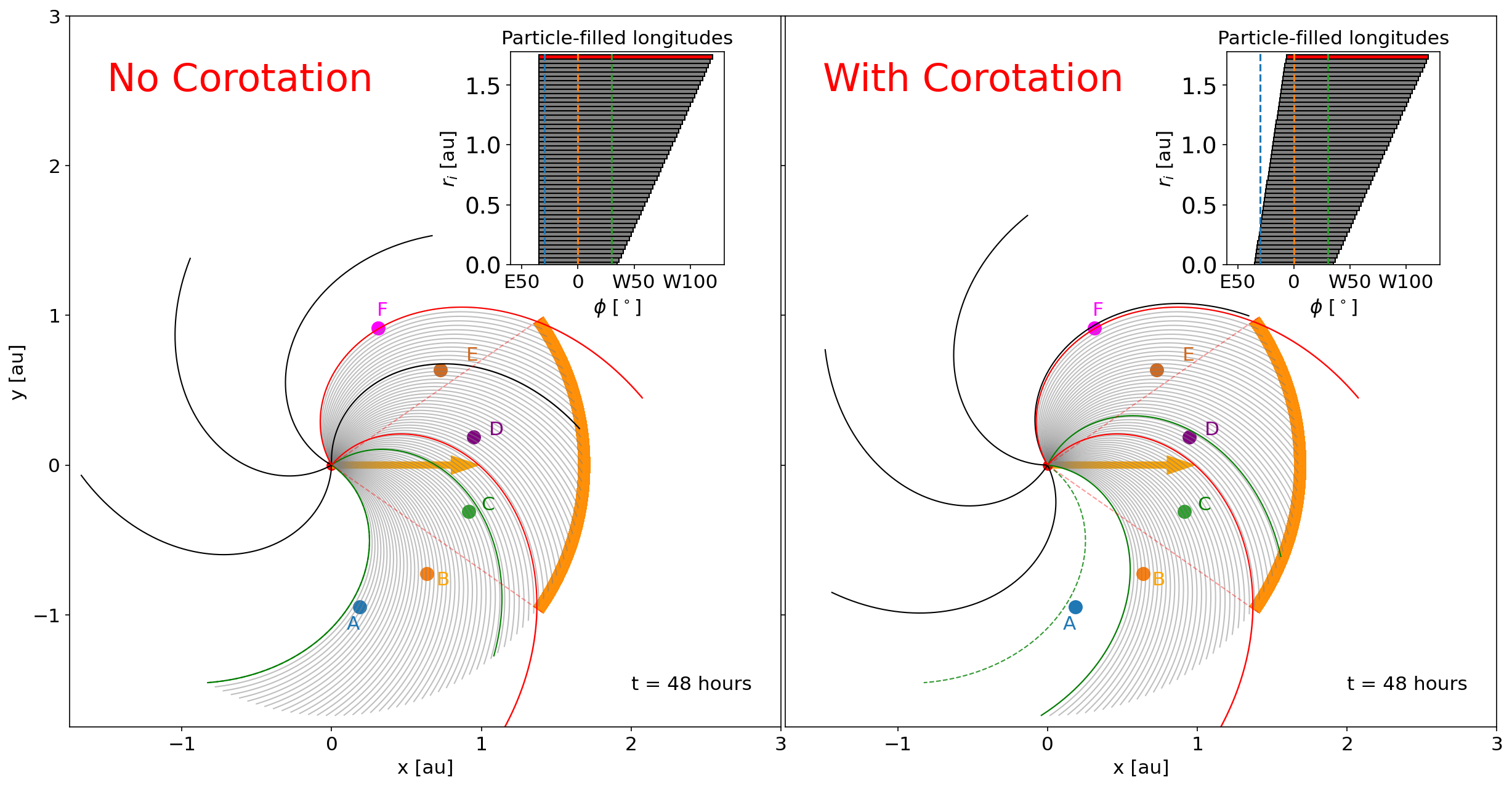}
    \caption{Schematic showing the geometry of the shock, observers, and particle-filled flux tubes, for the cases with (right panel) and without (left panel) corotation after 48 hours of shock propagation. Here x and y are heliocentric cartesian coordinates in the heliographic equatorial plane. 
    Observers A-F are denoted by the coloured circles. The shock's projection onto the plane is displayed here as the orange shaded segments. The grey IMF lines represent the range of particle-filled flux tubes. The inset shows the range of longitudes that are filled with particles. The solid red curved lines show the IMF lines that are currently connected to the edges of shock. The solid green curved lines show the magnetic flux tubes at the edges of the shock at the initial time. In the corotation case these have rotated with respect to their initial location, shown by the dashed green line. }
    \label{movie-still}
\end{figure*}

\begin{figure*}[htb]
    \centering
    \includegraphics[width = 0.7\linewidth, keepaspectratio = true]{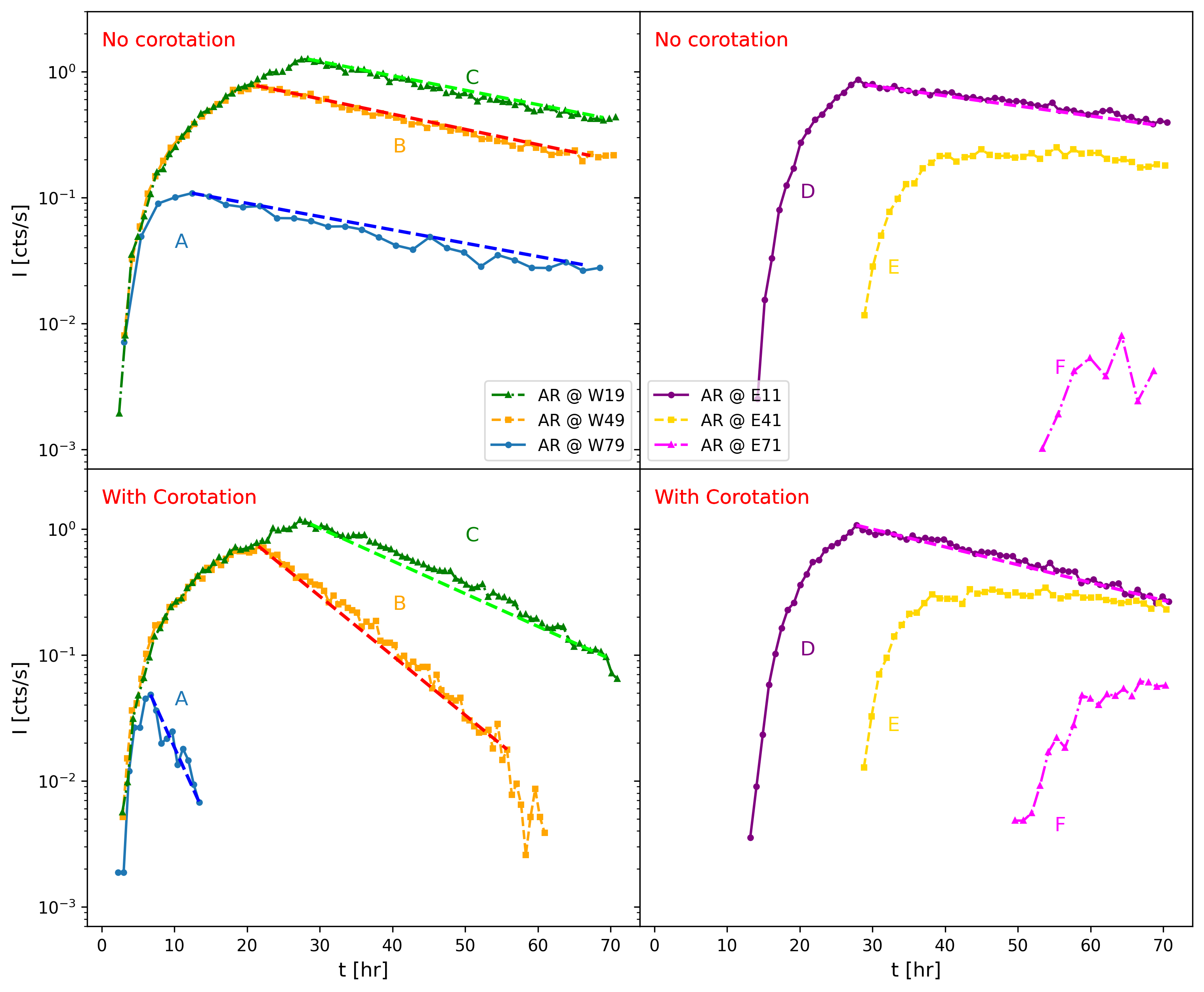}
    \caption{Intensity profiles for the six observers in Figure \ref{movie-still} without (top panels) and with (bottom panels) corotation for scattering conditions described by $\lambda = 0.1$ au. The dashed lines display the exponential decay fit for observers A-D. The shock reaches the 1 au distance at $t \sim 28$ hr.}
    \label{4-panel-obs-0_1au}
\end{figure*}


  
        

   
        


\section{Simulations}\label{sec:sims}

We use a 3D full-orbit test particle code \citep{Marsh_2013}, modified to describe a temporally extended injection of particles by a shock-like source. We use a model of an outward propagating shock, similar to that used by \cite{heras_1994} and \cite{Kal&Wib_1997}. A full description of the model is given in a companion paper \citep{hutch_2022-in_prep}. Here we summarise its main features. The shock-like source is a concentrically expanding front of fixed angular width.
Shock acceleration and downstream features are not modelled and the front only acts as a particle injector.

Particles are injected at time $t_{inj}$ at radial distance $r_{inj} = r_0 + v_{sh} \: t_{inj}$, where $r_0$ is the shock position at $t = 0$, and $v_{sh}$ is its velocity, assumed constant. Particles are injected uniformly across the shock front in both longitude and latitude. The number of particles injected by the shock at distance $r$ (radial injection profile, $R(r)$) is constant with $r$ for $r < r_{max}$, where we assume injection stops. 

In our simulations we follow a 5 MeV mono-energetic proton population, consisting of $N_p = 10^6$ particles. The particle crossing times at 1 au are collected to form intensity-time profiles at energy $>$1 MeV. The parameters of the shock front are as follows: shock speed $v_{sh} = 1500$ km s$^{-1}$, longitudinal and latitudinal width of the shock, $w_{\phi}$ = $w_{\delta}$ = $70^\circ$. The shock nose is located at heliolongitude $\phi_{nose}$ = $0^\circ$ and heliolatitude $\delta_{nose}$ = $15^\circ$. Injection at the shock ends at $ t = 48 $ hr, corresponding to $r_{max} = 1.73$ au, and we propagate SEPs until $t = 72$ hr.



Particles propagate in a unipolar Parker spiral IMF, a valid assumption when the SEP source region is far from the heliospheric current sheet, with constant solar wind speed $v_{sw} = 500$ km/s.
The shock does not disturb the Parker spiral and is not a magnetic enhancement. After injection the shock is transparent to particles. We consider the influence of IMF magnetic turbulence by including parallel scattering, parameterised by the mean free path, $\lambda$ \citep{Marsh_2013,dalla_2020}. Unless otherwise stated, we use $\lambda = 0.1$ au. The model does not include any perpendicular diffusion associated with turbulence, allowing us to investigate the corotation effects on SEP observables isolated from effects caused by cross-field diffusion \citep{He&Wan_2017}. 

In the test-particle code, corotation is accounted for by including the solar wind electric field, $\mathbf{E}$, into the equation of motion of the particle \citep{dalla_2013,Marsh_2013}.
The resulting electric field drift is a corotation drift \citep[see Equations 7-9 of][]{dalla_2013}, moving the particle's guiding centre by $\sim 14.2^\circ$ per day  (sidereal rotation rate, $\Omega = 2.86\times 10^{-6}$ rad/s) in the direction of solar rotation.
Setting $\mathbf{E} = 0$ in the test particle code switches off the corotation drift and allows one to study SEP propagation when corotation is neglected.  


\section{Intensity profiles at different observers}\label{sec:int_obs}

We study time-intensity profiles at six observers (labelled A - F, see Figure \ref{movie-still}). All observers are located at the same latitude as the shock nose ($\delta = 15^\circ$). We define the longitudinal separation, $\Delta \phi$, between the source Active Region (AR) and the observer's magnetic footpoint as,
\begin{equation}
    \Delta \phi = \phi_{AR} - \phi_{ftpt}
    \label{dphi}
\end{equation}
where $\phi_{AR}$ is the AR longitudinal position and $\phi_{ftpt}$ is the longitude of the observer footpoint. AR locations and corresponding $\Delta \phi $ values are given in Table \ref{event_tau_table}. A Parker spiral magnetic connection is assumed when calculating $\phi_{ftpt}$.

Figure \ref{movie-still} shows a schematic of the shock, observers and particle-filled magnetic flux tubes when corotation is excluded (\textit{left}) and included (\textit{right}) at $t = 48$ hr. The grey lines show the range of particle-filled flux tubes (i.e. IMF lines connected to the shock at $t \: \leq 48$ hr). The insets shows the range of longitudes filled with particles versus shock height. 

Figure \ref{4-panel-obs-0_1au} shows intensity profiles for observers A - F, without and with corotation (top and bottom rows respectively). Intensity profiles are obtained by collecting counts within $10^\circ \times 10^\circ$ tiles on the 1 au sphere. Features of the profiles such as the onset time and peak time relate to the establishment/loss of connection to the shock and its arrival at the spacecraft, as was noted in a number of previous studies \citep[e.g.,][]{heras_1994}. Here the inclusion/non-inclusion of corotation has a very significant effect on intensity profiles, in particular for observers A-C, which see the source AR as western.

\begin{figure}
    \centering
    \includegraphics[width = 0.95\linewidth, keepaspectratio = true]{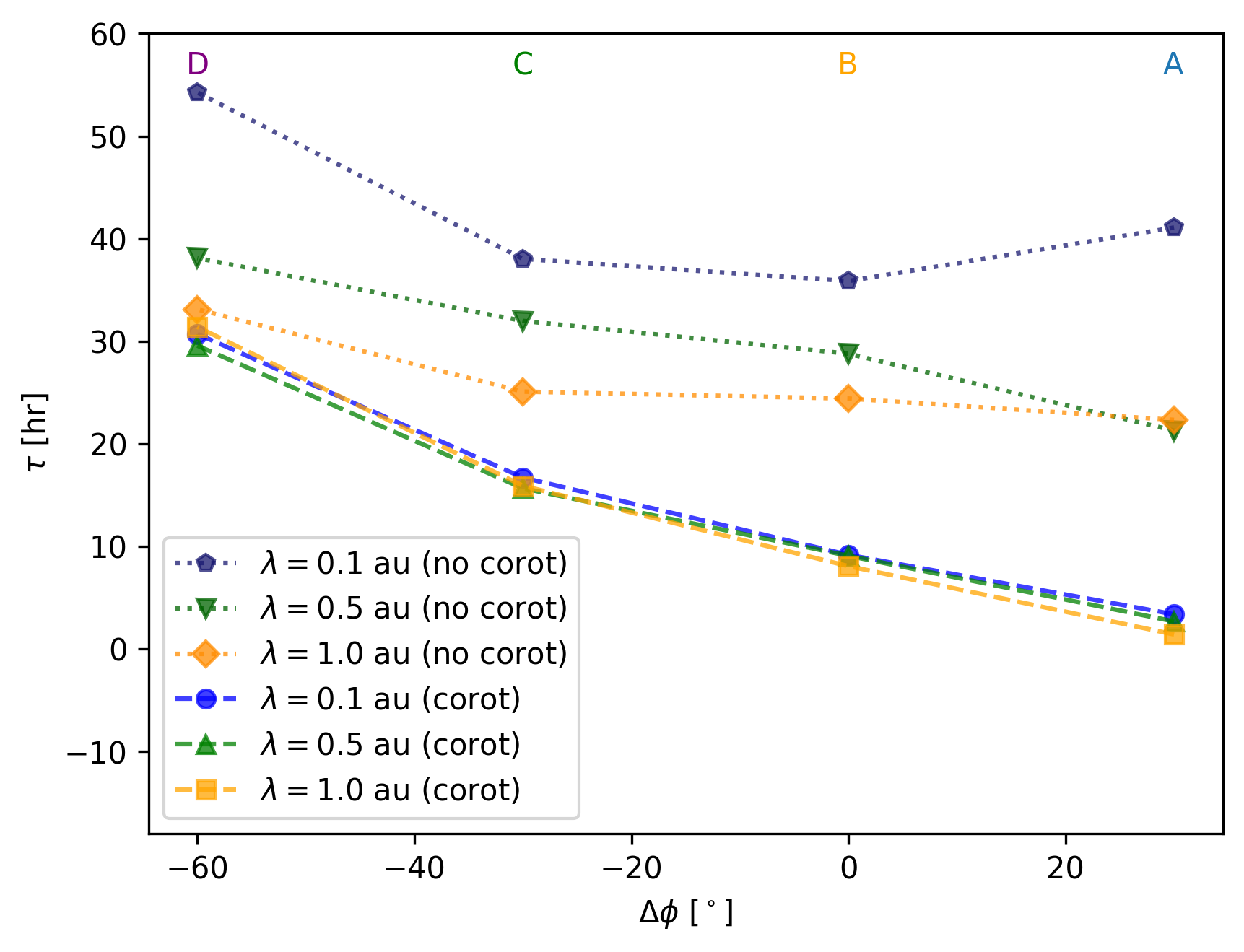}
    \caption{Decay time constant ($\tau$) for four of our observers with (dashed lines) and without (dotted lines) corotation and scattering conditions in the range $\lambda = 0.1$ au to $\lambda = 1.0$ au.}
    \label{tau_plt}
\end{figure}

Comparing top and bottom rows in Figure \ref{4-panel-obs-0_1au}, corotation has two main effects on the intensity profiles. Firstly, because the flux tubes are swept westward, the western events are cut short (observers A, B and C). For example, with corotation the duration for observer A drops significantly from $\sim 65$ hr to $\sim 12$ hr. Secondly corotation increases the intensity during eastern events (observers D, E and F).
For the observers located directly in the path of the shock (C and D) the effects of corotation are mainly seen after shock passage  ($t \sim $ 28 hr).
We quantify the effect of corotation on profiles for observers A-D by determining the decay time constant, $\tau$, by fitting an exponential between the peak intensity and a second point near the end of the profile, chosen by eye to avoid regions that fluctuate due to low counts. The decay time constants are given in Table \ref{event_tau_table} for the simulations in Figure \ref{4-panel-obs-0_1au}, and are plotted versus $\Delta \phi$ in Figure \ref{tau_plt}, where data points corresponding to $\lambda = 0.5$ and $1.0$ au are also shown, for simulations with (dashed lines) and without (dotted lines) corotation. There are no data points for observers E and F as there are no clearly defined decay phases in the intensity profiles. Figure \ref{tau_plt} shows a systematic shift to low $\tau$ for simulations with corotation, corresponding to faster decay phases for all observers. When corotation is included there is little dependence on the scattering conditions as corotation dominates the decay phase. When corotation is neglected $\tau$ is a measure of the degree of scattering, with smaller $\lambda$ leading to extended decay phases. 



\begin{table}[]
    \centering
    \begin{tabular}{ c c c c c }
            \hline
            {Observer} & AR location & {$\Delta \phi$ [$^{\circ}$]} & {$\tau_{\mbox{\tiny{no\_corot}}}$ [hr]} & {$\tau_{\mbox{\tiny{corot}}}$ [hr]} \\
           
            \hline 

         A & W79 &30 & 41.1 & 3.4 \\
         B & W49 &  0 & 35.9 & 9.2 \\
         C & W19 &-30 & 38.0 & 16.7 \\
         D & E11 & -60 & 54.3 & 30.7 \\
         E & E41 &-90 & - & - \\
         F & E71 & -120 & - & - \\
         
         \hline
    \end{tabular}
    \caption{Geometry of A-F observers and associated decay time constant of 5 MeV intensity profiles for $\lambda = 0.1$ au. Columns are from left to right: observer name, AR location with respect to the observer, $\Delta \phi$ as defined by Equation (\ref{dphi}), decay time constant without corotation $\tau_{\mbox{no\_corot}}$, decay time constant with corotation $\tau_{\mbox{corot}}$ for $\lambda = 0.1$ au. }
    \label{event_tau_table}
\end{table}

\begin{figure}
    \centering
    \includegraphics[width = 0.95\linewidth, keepaspectratio = true]{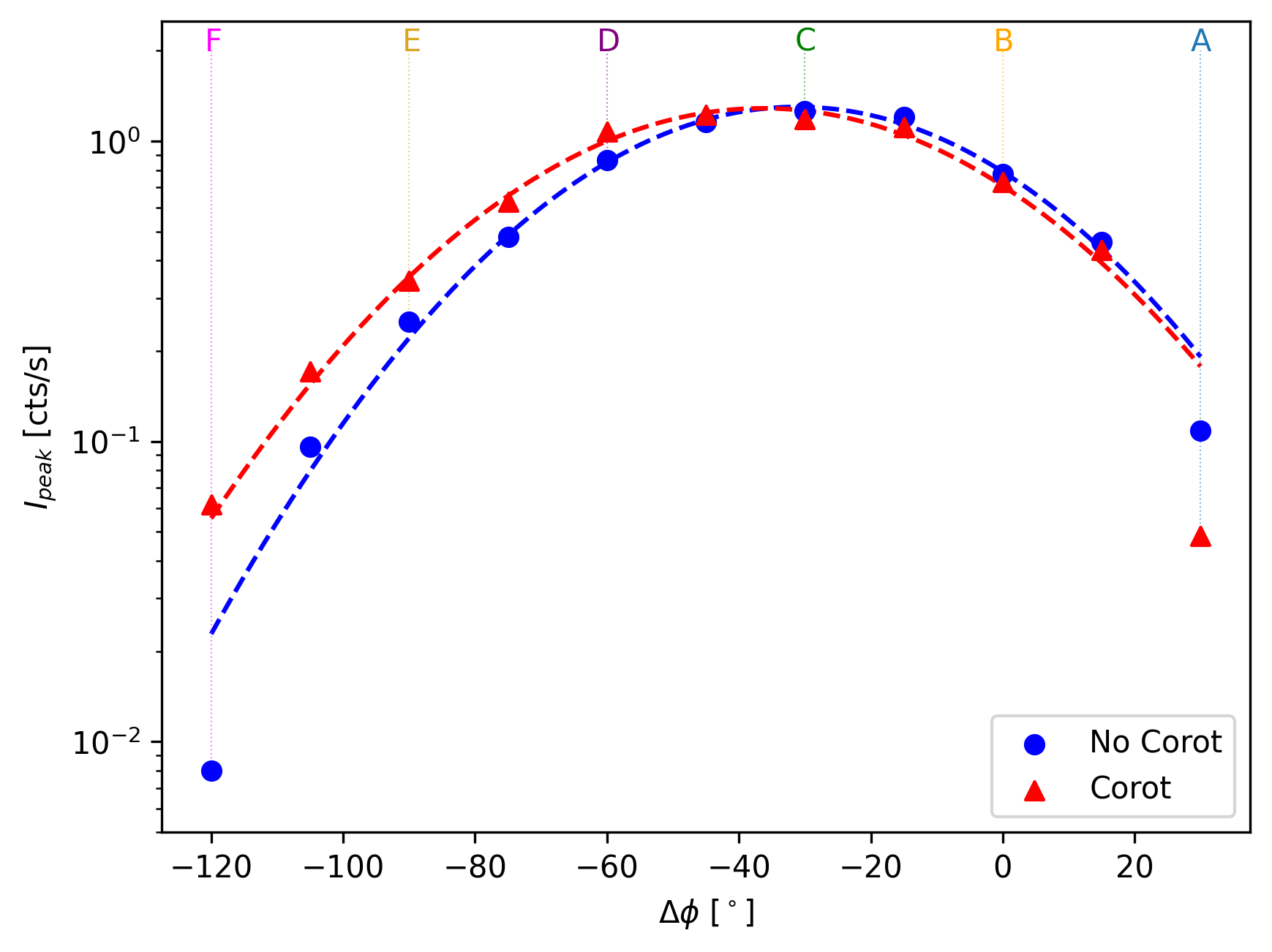}
    \caption{Peak intensity versus $\Delta \phi$ for the six observers A-F with (red triangles) and without (blue circles) the effects of corotation for scattering conditions described by $\lambda = 0.1$ au. The points are fit with Gaussian functions shown as the dashed lines.}
    \label{max_int}
\end{figure}

The peak intensity, $I_{peak}$, is plotted versus $\Delta \phi$ in Figure \ref{max_int}. $I_{peak}$ is largest for observers C and D, which are directly in the path of the shock. We fitted both sets of points with a Gaussian centred at $\Delta \phi = \phi_0$. Figure \ref{max_int} shows that $\phi_0$ is shifted with respect to the well-connected location ($\Delta \phi = 0$), with $\phi_0 = -36.4^\circ$ for the corotation fit and $\phi_0 = -31.2^\circ$ for no corotation. The corotation fit has a standard deviation $\sigma = 33.4^\circ$ and the no corotation $\sigma = 31.3^\circ$. This east-west asymmetry has been noted by several authors \citep{Ding_2022,Lario_2013,He&Wan_2017}. Figure \ref{max_int} shows that corotation enhances the asymmetry. 

\begin{figure}
    \centering
    \includegraphics[width = 0.8\linewidth, keepaspectratio = true]{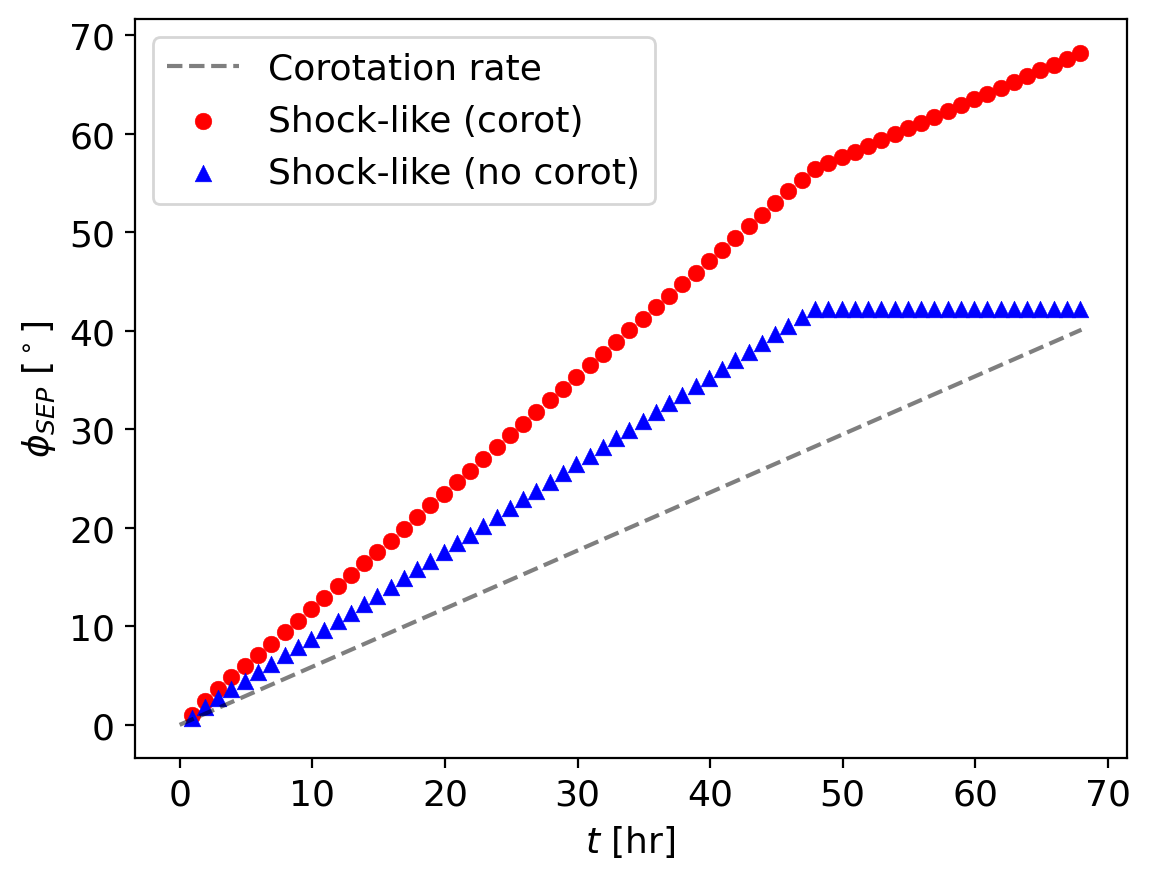}
    \caption{Median longitudinal position of the SEP population, $\phi_{SEP}$, for the shock-like injection with corotation (red circles), without corotation (blue triangles). The grey dashed line shows the corotation of the flux tube connected to the source region at $t = 0$. } 
    \label{median_part_long}
\end{figure}

Figure \ref{median_part_long} shows the median longitude, $\phi_{SEP}$, of all the test particles in our simulation versus time. The geometry of the shock connection to the observer already naturally produces a westward shift of $\phi_{SEP}$ with time, as shown by the blue triangles \citep{Ding_2022}. When corotation is taken into account, the latter effect becomes more pronounced (red points), resulting in the larger $\sigma$ for the corotation points in Figure \ref{max_int}. We note that the discontinuity at $t = 48$ hr is due to injection by the shock ending at this time.


\section{Discussion and conclusions}\label{sec:disc}

In this work we simulated particle injection from a shock-like source using 3D test particle simulations and compared intensity profiles over a wide range of observer longitudes with and without corotation. 

The main conclusions of our work are as follows:

\begin{enumerate}
    \item Corotation of particle-filled flux tubes has a strong effect on SEP intensity profiles for the case of time-extended acceleration at a propagating CME-shock (Figure \ref{4-panel-obs-0_1au}). Its main influence is on the decay phase of the event: e.g. it reduces the decay time constant compared to the case when corotation is not included. The strongest corotation effects are on observers that see the source AR in the west: both the event duration and decay time constant are significantly reduced (Table \ref{event_tau_table} and Figure \ref{tau_plt}).
    
    \item Corotation enhances maximum intensities during eastern events and makes the east-west asymmetry in peak intensity versus $\Delta \phi$ stronger (Figure \ref{max_int}).
    
    \item Deriving intensity profiles without including the effects of corotation (by solving particle transport equation in the corotating frame or using a 1D approach that models propagation along a single flux tube) artificially extends the decay phase, especially for western events.
    
    \item Varying the scattering mean free path between $\lambda = 0.1$ and $1.0$ au has very little influence on the decay phase (negligible difference in the value of $\tau$) when corotation is included, indicating that corotation is a dominant process during the decay phase of SEP events.
\end{enumerate}

Our simulations show that, unlike previously thought \citep[e.g.][]{Kal&Wib_1997,Lario_1998}, corotation is a key influence on the decay phase of SEP events and it also affects the peak intensity phase (excluding possible Energetic Storm Particle (ESP) enhancements). Within the large variability in the properties of SEP events, there is some indication that corotation plays a role. \cite{Dalla_2003} analysed the duration of 52 events and showed that there is a tendency for eastern events to have longer durations compared to western ones. The longitudinal dependence of the SEP spectral index, first reported by \cite{van_H_1975}, can be explained by corotation effects: eastern events take a long time to corotate to an Earth observer and for this reason their spectral index is larger as the high energy particles have escaped the inner heliosphere by the time the flux tube corotates over the observer. 

In the work presented here we have used a uniform rate of injection from the shock with radial distance and longitude/latitude across the shock. \cite{hutch_2022-in_prep} have also considered two other radial injection functions: we verified that corotation plays an important role regardless of the details of the injection function. Similarly, they showed that varying the spatial profile of injection along the shock front has a minor effect on the intensity profiles.

Our shock model is intended as only an initial approximation, given that the present work is the first analysis of a moving shock source within 3D test particle simulations. In particular downstream features, such as a flux rope and non-Parker field lines, are not described. We note that MHD simulations of shock propagation show that the magnetic field lines in the downstream can connect back to the shock. In 3D the magnetic field lines are known to wrap over/under the ejecta \citep{Lario_1998}: we expect that there are many cases where an observer behind the shock would be connected to it. Our study applies directly to these cases. With a more accurate model of the downstream region intensities may differ from those in Figure \ref{4-panel-obs-0_1au} behind the shock, however they would still be influenced by corotation of the magnetic flux tubes. We hope to include such a model in future studies.

We note that when constructing time-intensity profiles as shown in Figure \ref{4-panel-obs-0_1au}, we used all particles ($> 1$ MeV), although some of them have an energy lower than the initial $5$ MeV, due to adiabatic deceleration. We have determined intensity profiles for particles in the energy range 4.5-5.0 MeV (not shown), which show very similar behaviour to those in Figure \ref{4-panel-obs-0_1au}, displaying even smaller decay time constants in the corotation case. During an SEP event, particles over a range of energies will be injected and those produced with initial energy higher than 5 MeV will decelerate into the latter energy band.

A number of studies have used focussed transport or Fokker-Planck equations to determine intensity profiles after injections from a CME-driven shock-like source \citep[e.g.][]{He_2015,Wang_2012}. In contrast to our results their intensity profiles look very similar across a range of longitudinal positions for the observer, and this may be due to neglecting corotation. 



We have derived other observables such as the onset time and time of peak intensity from our simulations. However, corotation does not have a significant effect on these quantities as they are primarily determined by geometric factors, such as times of observer-shock connection and time of shock passage at the observer. 



In the present study the role of corotation has been investigated without including the effect of possible turbulence-induced perpendicular diffusion, associated with magnetic field line meandering. Traditionally in the interpretation of gradual SEP events this effect has not been included \citep[e.g][]{Reames_1997,Lario_1998, bain_2016}, although other researchers have emphasised its importance \citep[e.g.][]{Wang_2012,He_2015, He&Wan_2017}, in models that do not include corotation. We aim to include this effect in future work. One would expect perpendicular diffusion to produce earlier SEP onsets for eastern events and to help SEP intensities reach similar values at far away locations faster. 

Our study has used a simplified model of injection at a shock to highlight the effects of corotation and geometry of  observer-shock connection. In future work we hope to use a more sophisticated description of shock acceleration to model specific SEP events. 
SEP intensity profiles may be influenced by complex coronal and interplanetary magnetic field configurations and solar wind structures, which were not considered in the present study. A full analysis of these effects requires detailed modelling event by event.

\bibliographystyle{aa}
\bibliography{corot}

\begin{acknowledgements}
A. Hutchinson, S. Dalla and T. Laitinen acknowledge support from the UK Science and Technology Facilities Council (STFC), through a Doctoral Training grant - ST/T506011/1 and grants ST/R000425/1 and ST/V000934/1. 
C.O.G. Waterfall and S. Dalla acknowledge support from NERC grant NE/V002864/1.

This work was performed using resources provided by the Cambridge Service for Data Driven Discovery (CSD3) operated by the University of Cambridge Research Computing Service (www.csd3.cam.ac.uk), provided by Dell EMC and Intel using Tier-2 funding from the Engineering and Physical Sciences Research Council (capital grant EP/P020259/1), and DiRAC funding from the Science and Technology Facilities Council (www.dirac.ac.uk).
\end{acknowledgements}

\end{document}